# Magnetic Shape Memory Microactuator

E. Kalimullina[1], A. Kamantsev[*,1,2], V. Koledov[1,2], V. Shavrov[1], V. Nizhankovskii[2], A. Irzhak[3,4], F. Albertini[5], S. Fabbrici[5,6], P. Ranzieri[5], P. Ari-Gur[7]

[1] Kotelnikov Institute of Radio-engineering and Electronics of RAS, 125009 Moscow, Russia
[2] International Laboratory of High Magnetic Fields and Low Temperatures, 53-421 Wroclaw, Poland
[3] National University of Science and Technology ''MISiS'', 119049 Moscow, Russia
[4] Institute of Microelectronics Technology and High-Purity Materials of RAS, 142432 Chernogolovka, Russia
[5] IMEM-CNR, 43124 Parma, Italy
[6] MIST-ER laboratory, 40129 Bologna, Italy
[7] Western Michigan University, MI 49008 Kalamazoo, USA

Bimetallic composite nanotweezers based on $Ti_2NiCu$ alloy with shape memory effect (SME) have recently demonstrated the ability to manipulate real nano-objects, such as nanotubes, and bionanoparticles when heated to 40-60 °C by laser radiation. The possibility of developing nanotweezers operating at constant temperature is of particular importance mainly for the manipulation of biological objects. In this work, a microactuator was produced using a composite bilayer made of a layer of rapidly quenched $Ni_{53}Mn_{24}Ga_{23}$ ferromagnetic shape memory Heusler alloy and an elastic layer of Pt. The size of the microactuator is $25 \times 2.3 \times 1.7$ μm$^3$. A controlled bending deformation of the actuator of 1.2 %, with a deflection of the end of the actuator higher than 2 μm was obtained by applying a magnetic field of 8 T at T = 62 °C. The possibility of the development of new technologies for magnetic-field-controlled nanotools operating at a constant temperature using the new multifunction magnetic shape memory alloys will be discussed.

**1 Introduction** Heusler alloys attract great interest due to the combination of ferromagnetism and thermoelastic martensitic transition (MT), which is accompanied by shape memory effect (SME) [1]. Recently, application of the technology of selective ion etching allowed for the creation of two-layer composite actuators and tools based on rapidly quenched nonmagnetic alloys with SME, such as $Ti_2NiCu$ [2]. These composite actuators can change their shape reversely and produce mechanical work using only "one-way" SME of the alloy [3]. Currently in the field of manipulation and manufacturing at the nanoscale, there is an urgent need to develop new functional materials in order to fill the gap between the dimensions of modern MEMS and real size of nano-objects to be manipulated. Recently the operation of nanotweezers using layered composites based on alloy with SME driven by thermal actuation has been demonstrated [2-4]. The operation of the $Ti_2NiCu/Pt$ composite actuator driven by heating was demonstrated, with an overall volume of the actuator of less than 1 μm$^3$ and thickness of active layer of the $T_2NiCu$ alloy being as small as 170 nm [2]. This opens up the possibility of developing technology for the production of micro-sized magnetic-field-controlled tools and devices based on Heusler alloys. The idea of the present work is to use the phenomenon of magnetic-field-controlled SME in Heusler alloy $Ni_2MnGa$ to design a composite magnetic microactuator with SME operating at constant temperature.

**2 Samples** The microactuators were prepared by standard Focused-Ion-Beam and Chemical-Vapor-Deposition (FIB-CVD) processes in FEI Strata 201 FIB device (see Fig. 1) [5]. An alloy with a composition of $Ni_{53}Mn_{24}Ga_{23}$ was chosen for the experiments due to the convenient temperature range of MT and its small hysteresis. Melt-spun ribbons of 30-μm-thickness with nominal composition of $Ni_{53}Mn_{24}Ga_{23}$ were prepared by melt spinning. Some pieces of ribbons were annealed in vacuum at 800°C for 5 and 72 hours. At room temperature all samples were in the ferromagnetic and martensitic state. Their Curie point ($T_C$) as well as their start and finish temperatures of the austenite-martensite and martensite-austenite transformations ($M_s$, $M_f$, $A_s$ and $A_f$, respectively) were determined by using DSC. For the sample, annealed for 72 hours, they were $M_s$ = 49.5, $M_f$ = 41.2, $A_s$ = 50.4, $A_f$ = 60.7, and $T_C$ = 72.5 °C. More details on the preparation and other parameters of the ribbons are given elsewhere [5-8]. The principle of actuation of composites with SME requires that at a first stage the active layer of SME alloy is pre-strained Then a passive elastic layer of Pt has to be deposited on the active one (Fig. 1a) [1]. Then the composite is nanofabricated by ion beam cutting from the ribbon surface. Cantilever type micro-actuators of size $25 \times 2.3 \times 1.7$

µm$^3$ are then prepared and moved by Omniprobe micromanipulator from the surface of the ribbon and then attached by CVD process to Si substrates (Fig. 1b).

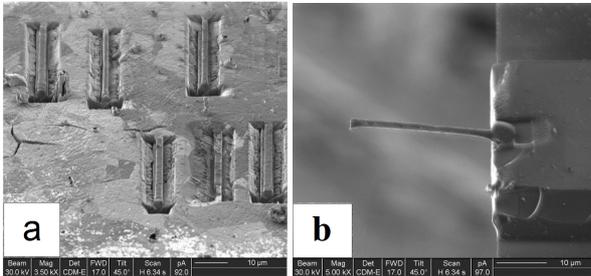

**Figure 1** Preparation of microactuators on the base of $Ni_{53}Mn_{24}Ga_{23}$/Pt composites with SME: (a) preforms of composites on the surface of prestrained $Ni_{53}Mn_{24}Ga_{23}$ melt spun ribbon; (b) a composite microactuator attached to Si substrate.

**3 Experimental description** The experiments on magnetic actuation have been carried out in 8 T Bitter coil at the International laboratory of strong magnetic fields and low temperatures, Wroclaw, Poland. For the observation of the magnetic-field-controlled giant bending strain of the microactuators (see Fig. 2) a non-magnetic optical microscope with vacuum thermostat for temperature control was purpose-built.

The effect of magnetic-field-controlled MT in ferromagnetic alloys with SME is based on the temperature shift of MT in magnetic field (see Fig. 3a) due to different magnetization values of the martensitic and austenitic phase of the alloy. So by choosing the temperature near Mf one can cause the reversible MT turning on the sufficiently strong magnetic field. The sensitivity of MT of the alloy $Ni_{53}Mn_{24}Ga_{23}$ to magnetic field is near 1 K/T [8]. Therefore, an almost complete reversible MT accompanied by giant bending strain of the composite can been observed in a field of 8 T (Fig. 3b). The relative bending strain has been determined from the microscopic images of the composites using the formula: ε = h/2R, where h – thickness of the composite, R – radius of the composite curvature. The controlled relative bending is not less than 1.2 %, the controlled stroke of the tip of composite is up to 2 µ. The composite is bent in zero field and straitened almost completely at $\mu_0H = 8$ T due to the interaction between the elastic layer of Pt and prestrained active SME layer experiencing magnetic-field-induced MT.

Recently the new metamagnetic Heusler functional alloys Ni-Mn-In-Co and Ni-Mn-Ga-Co have been a subject of intensive investigation [9, 10]. The martensitic phase of these alloys is almost nonmagnetic; the application of a magnetic field induces a shift of MT to lower temperatures with a sensitivity of up to 10 K/T [9]. Further work is still needed to diminish the hysteresis of MT in order to use lower fields (H < 2 T) and to optimize the temperature range of the actuation. The new technology of magnetic field controlled actuation of living microobjects using technological and cheap permanent magnets could help to resolve important medical and biological tasks, such as single cell manipulation, microsurgery operations at micro- and submicrometer level. The new visualization techniques on submicron scale should also be developed.

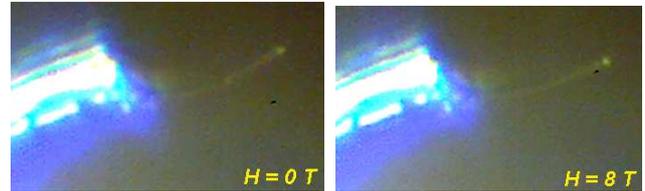

**Figure 2** Giant magnetic-field-controlled bending strain of composite microactuator in Bitter coil at constant temperature The composite is bent in zero field and straitened almost completely at $\mu_0H = 8$ T due to interaction of the elastic and prestrained active layers with SME.

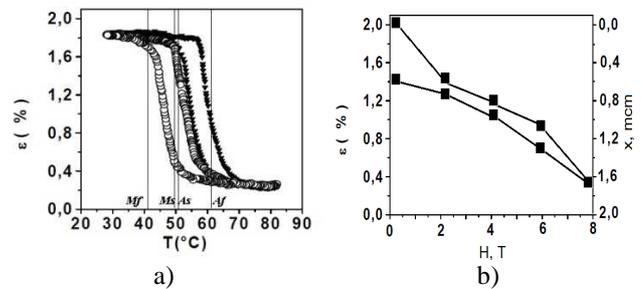

**Figure 3** Magnetic-field-controlled SME in $Ni_{53}Mn_{24}Ga_{23}$ alloy: (a) relative strain of $Ni_{53}Mn_{24}Ga_{23}$ melt spun ribbon versus temperature at $\mu_0H = 0$ T (○) with transition temperatures and at $\mu_0H = 6$ T (▼) [8]. (b) Magnetic-field-controlled bending strain ε and stroke x of $Ni_{53}Mn_{24}Ga_{23}$/Pt composite microactuator at constant temperature T = 62 °C.

**4 Conclusions** 1) The microactuator was created by standard FIB-CVD process starting from a composite bilayer made of one layer of rapidly quenched ferromagnetic shape memory alloy based on Ni-Mn-Ga and an elastic layer of Pt. The size of the microactuator is 25×2.3×1.7 µm$^3$. 2) A controlled bending strain of the actuator of 1.2 % was obtained and the stroke of the actuator was not less than 1.6 µm in a magnetic field $\mu_0H = 8$ T at a constant temperature T = 62 °C.

**Acknowledgements** Authors are grateful to Professors V. Pushin, S. Belyaev for discussions. The work was supported by RFBR grants No. 12-08-01043-a, 12-07-00656-a, 12-08-31340-мол_a, Russian Ministry of Science and Education Agreement No 8571, RAS-CNR Joint Research Program, and by the 2007-2013 FESR Operative program of the Emilia Romagna Region (Activity l.1.1). The equipment of CKP MIPT and REC "Nanotechnology" of MIPT was used in this work. Also, award No. RUP1-7028-MO-11 of the US Civilian Research & Development Foundation (CRDF) is acknowledged.